\documentclass{elsart5}
\usepackage{graphicx}
\usepackage{pslatex}
\usepackage{amssymb,amsmath}

\begin{document}

\begin{frontmatter}

\title{Critical properties of the XXZ model with long-range interactions on the double chain}

\author[aff1]{J. P. de Lima\corauthref{cor1}}
\ead{pimentel@ufpi.br}
\corauth[cor1]{}
\author[aff2]{L. L. Gon\c calves}
\address[aff1]{Departamento de F�'isica, Universidade Federal do Piau\'i, Campus Min. Petr\^onio Portela, 64049-550 Teresina, Piau\'i, Brazil}
\address[aff2]{Departamento de Engenharia Metal\'urgica e de Materiais, Universidade Federal do Cear\'a, Campus do Pici, 60455-760 Fortaleza, Cear\'a,  Brazil }
\received{12 June 2005}
\revised{13 June 2005}
\accepted{14 June 2005}


\begin{abstract}
 The $XXZ$ model $(s = 1/2)$ in a transverse field on a double chain with a uniform
long-range interaction among the $z$ components of the spins is
considered. The nearest-neighbour interactions are restricted to the
components in the $xy$ plane and to the spins within the same chain
leg, such that the Hamiltonian is given by $H = -\sum_{m=1}^{2}
J_{m} \sum_{j=1}^{N}(S_{m,j}^{x}S_{m, j+1}^{x} +
S_{m,j}^{y}S_{m,j+1}^{y}) - \frac{I}{N}\sum_{m,n=1}^{2}
\sum_{j,k=1}^{N}S_{m,j}^{z}S_{n,k}^{z}-h\sum_{m=1}^{2}
\sum_{j=1}^{N}S_{m,j}^{z}$, where $N$ is the number of sites of the
lattice and $m,n$ $( m,n = 1, 2)$ label the chain legs. The model is
solved exactly by introducing the Jordan-Wigner
 and integral Gaussian transformations, which map the Hamiltonian in a
 non-interacting fermion system and
corresponds to an extension of the model recently studied by the
authors for a single chain. The equation of state is obtained in
closed form, and the critical classical (at $T
> 0$) and quantum (at $T = 0$) behaviours can be determined exactly.
The quantum critical surface is determined in the space generated by
the transverse field and interaction parameters, and the crossover
lines separating the different critical regimes are also obtained.
It is also shown that, differently from the results obtained for the
single chain, the system can present multiple quantum transitions.

\end{abstract}


\end{frontmatter}

The critical behaviour of the XXZ model with long-range interactions among
the spins along the z direction, in a single chain, has been recently
considered by the authors \cite{goncalves2005}. In that paper they present a
comprehensive study of the quantum and classical transitions undergone by
the model. Although the model presents first- and second- order quantum and
classical transitions, which is a very important feature, no multiple
transitions are present. However, multiple quantum transitions can occur in
magnetic systems as it has been shown in the experimental results recently
obtained for ZrZn$_{2}$ \cite{uhlarz2004}. This result has been the main
motivation for studying spin models which are exactly soluble and can
present multiple quantum transitions. Therefore we will consider the XXZ
model $(s=1/2)$ in a transverse field on a double chain with long-range
uniform interaction among z components of spins. The nearest-neighbour
interactions are assumed isotropic, and restricted to the spin components in
the XY plane and to the spins within the same chain leg. It should be noted
that due to the isotropy of the nearest-neighbour interactions the total
magnetization along the $z$ direction is conserved. Assuming periodic
boundary conditions the Hamiltonian is given by
\begin{equation}
\begin{split}
H& =-\sum\limits_{m=1}^{2}J_{m}\sum\limits_{j=1}^{N}\left(
S_{m,j}^{x}S_{m,j+1}^{x}+S_{m,j}^{y}S_{m,j+1}^{y}\right) \\
& -\sum\limits_{m,n=1}^{2}\frac{I}{N}\sum%
\limits_{j,k=1}^{N}S_{m,j}^{z}S_{n,k}^{z}-h\sum\limits_{m=1}^{2}\sum%
\limits_{j=1}^{N}S_{m,j}^{z}
\end{split}
\label{hamiltonian}
\end{equation}
where $N$ is the number of sites of one leg and $m$ ( $m=1,2$ )
labels the chain legs. By introducing the Jordan-Wigner transformation \cite%
{jordan-wigner1928} and a Gaussian transformation, we can write the
partition function as
\begin{eqnarray}
Z_{N} &=&\left( \frac{N}{2\pi }\right) ^{1/2}\int_{-\infty }^{\infty }dx\exp %
\left[ -\frac{N}{2}x^{2}\right] \times Tr\exp [-\overline{H}]\times \\
&&\exp \left[ -N\sum (\beta h+\sqrt{2\beta I})\right] ,  \notag
\end{eqnarray}%
where $\beta =1/k_{B}T$ , the effective Hamiltonian $\overline{H}$ is given
by
\begin{equation}
\overline{H}\equiv \sum_{m,j}\beta J_{m}(c_{m,j}^{\dag
}c_{m,j+1}+c_{m,j+1}^{\dag }c_{m,j})+\sum_{m,j}(\text{ }\beta h+\sqrt{2\beta
I})c_{m,j}^{\dag }c_{m,j},
\end{equation}%
and it has been used the fact that the long-range interaction commutes with $%
H.$ Then, introducing the spatial Fourier transform, the Hamiltonian $\overline{H}$ , apart
from a constant, can be written in diagonal form with energy excitations $\overline{E}%
_{k_{m}}=\beta J_{m}\cos (k_{m})+\beta h+\sqrt{2\beta I}x.$ Then the
partition function can be recast in the form
\begin{equation}
Z_{N}=\left( \frac{N}{2\pi }\right) ^{1/2}\int_{-\infty }^{\infty }dx\exp %
\left[ -\frac{N}{2}x^{2}\right] \prod_{m,k}\left[ 1+\exp (-\overline{E}%
_{k_{m}})\right]  \label{partition}
\end{equation}%
and, in the thermodynamic limit, can be evaluated by the saddle-point
method. Following this procedure, the free energy functional can be written
in terms of the magnetization, $M^{z},$ and at $T=0,$ is given by
\begin{equation}
f=h+4IM^{z}(M^{z}+1)-\frac{1}{\pi }\left( J_{1}\sin {%
\varphi _{1}}+J_{2}\sin {\varphi _{2}}\right) -\frac{1}{\pi }\left( \varphi
_{1}+\varphi_{2}\right) ,  \label{func}
\end{equation}%
where $M^{z}$ is explicitly given by
\begin{equation}
M^{z}=\frac{\varphi _{1}+\varphi _{2}}{2\pi }-\frac{1}{2},  \label{mag}
\end{equation}%
with $\varphi _{m}=\arccos (-(h+4IM^{z})/J_{m}).$ It should be noted that for
$J_{1}=J_{2}$ the results obtained so far reproduce the known ones for the
single chain \cite{goncalves2005}. By minimizing the free energy functional,
given in eqs (\ref{func}) and (\ref{mag}), and the by using stability
condition
\begin{equation}
\frac{\partial ^{2}f}{\partial {(M^{z})}^{2}}\geq 0,
\end{equation}%
we can determine the equation of state and obtain the phase diagram at $T=0.$
Whithout loss of generality, we will restrict the analysis of the phase
diagram to the region $J_{2}/J_{1}\geqslant 1,$ which presents the main
features of the model. In order to charaterize the different quantum phases
displayed by the model we present in Fig. 1 the magnetization as a function
of the field for different values of the long-range interaction. As it can
be seen, for $I\leq 0,$ the transitions are of second order since the
magnetization is continuous at the transition points. On the hand, for $I>0$
the magnetization is discontinuous at the the transition points which is a
signature of a first order transition.

The quantum critical surface obtained is shown in Fig. 2 where, as already
mentioned, we have quantum phase transitions of first and second order. In
the region $I\leq 0,$ it can be shown that there are two critical surfaces
of second order quantum phase transition. These critical surfaces are given
explicitly by the expressions $h=J_{1}-2I\arccos \left( -J_{1}/J_{2}\right)
/\pi $, which corresponds to the magnetization $M^{z}=1/2\pi \arccos \left(
-J_{1}/J_{2}\right) ,$ and by $h=J_{2}-2I$, which corresponds to the
magnetization $M^{z}=1/2$. These two surfaces intercept when $J_{1}=J_{2}$
as it can be seen in the diagram shown in Fig.2. For $I>0$ the model
presents three lines corresponding to the quantum phase transitions of first
order, with a triple line which is also shown in Fig. 2. In this case, it is
not possible to obtain an explicit equation for the critical surface and it
has been determined numerically.

The interception of the critical surfaces of first and second orders occurs
at $I=0$, and corresponds to a bicritical line since it contains the known
bicritical point present for $J_{1}=J_{2}$ \ \cite%
{goncalves2005,goncalves2001}, when the results reduce to the ones for the
single chain.

Moreover, it should be noted that, for a given set of parameters, the system
presents multiple transitions as a function of the field, as it can be seen
in Fig. 2. In the second-order region, the multipĺe transitions can 
also be obtained from the explicit equation for the critical surface given above.

Finally, we would like to point out that, for $J_{1}=J_{2},$ the quantum
critical behaviour is identical to the one obtained for the model when we
have a single chain \cite{goncalves2005,goncalves2001}, and as already
mentioned we do not have multiple transitions.

\begin{figure}[b]
\begin{center}
\includegraphics[width=\columnwidth]{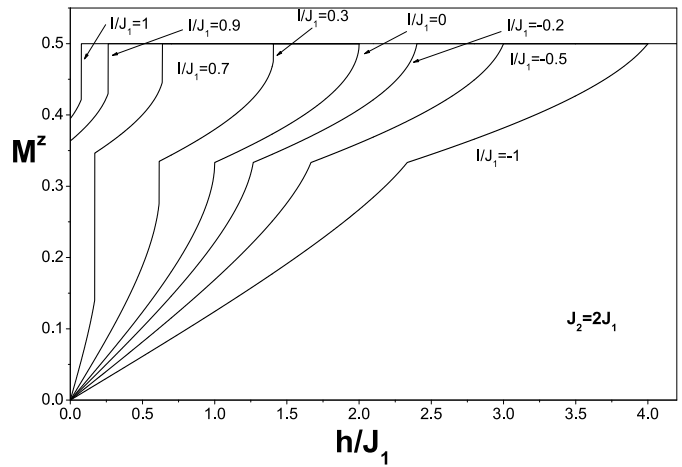}
\end{center}
\caption{Magnetization (at $T=0$) as a function of the field for
different values of the long-range interactions.} \label{fig-0}
\end{figure}

\begin{figure}[b]
\includegraphics[scale=0.35]{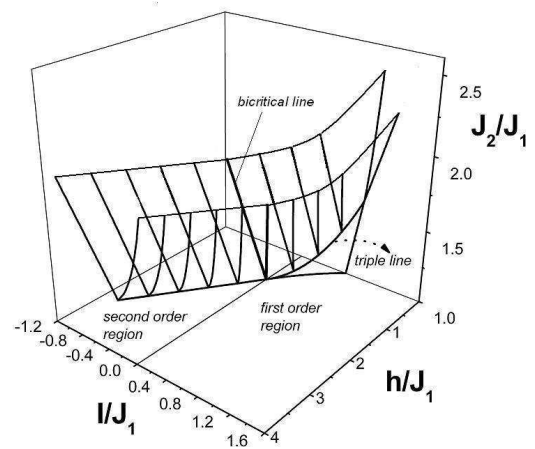}
\caption{Phase diagram of the model at $T=0$ and $J_{2}/J_{1}>1$.
} \label{fig-1}
\end{figure}

This work was partially financed by the Brazilian agencies CNPq and Capes.
The authors would like to thank Dr. A. P. Vieira for a critical reading of
the manuscript.

\end{document}